\documentclass[12pt,thmsa]{article}
\usepackage{amssymb}

\usepackage{sw20jart}

\input{tcilatex}
\input tcilatex
\begin{document}

\title{Extending the Lambda Calculus to Express Randomized and Quantumized
Algorithms}
\author{Philip Maymin\thanks{%
The author's email address is \texttt{pzmaymin@fas.harvard.edu}.} \\
Harvard University}
\date{December 31, 1996 }
\maketitle

\begin{abstract}
This paper introduces a formal metalanguage called the lambda-q calculus for
the specification of quantum programming languages. This metalanguage is an
extension of the lambda calculus, which provides a formal setting for the
specification of classical programming languages. As an intermediary step,
we introduce a formal metalanguage called the lambda-p calculus for the
specification of programming languages that allow true random number
generation. We demonstrate how selected randomized algorithms can be
programmed directly in the lambda-p calculus. We also demonstrate how
satisfiability can be efficiently solved in the lambda-q calculus.
\end{abstract}

\section{Introduction}

This paper presents three formal language calculi, in increasing order of
generality. The first one, the $\lambda $-calculus, is an old calculus for
expressing functions. It is the basis of the semantics for many functional
programming languages, including Scheme \cite{R4RS}. The second one, the $%
\lambda ^{p}$-calculus, is a new calculus introduced here for expressing
randomized functions. Randomized functions, instead of having a unique
output for each input, return a distribution of results from which we sample
once. The third one, the $\lambda ^{q}$-calculus, is a new calculus
introduced here for expressing quantumized functions. Quantumized functions
also return a distribution of results, called a \emph{superposition}, from
which we sample once, but $\lambda ^{q}$-terms have signs, and identical
terms with opposite signs are removed before sampling from the result. Thus,
superpositions can appear to shrink in size whereas distributions cannot.
The $\lambda ^{p}$-calculus is an extension of the $\lambda $-calculus. The $%
\lambda ^{q}$-calculus is an extension of the $\lambda ^{p}$-calculus. The $%
\lambda ^{q}$-calculus is the most general but it is best presented in
reference to the intermediary $\lambda ^{p}$-calculus.

Although much research has been done on the hardware of quantum computation
(c.f. \cite{deutsch 85}, \cite{deutsch 89}, \cite{simon}), none has focused
on formalizing the software. Quantum Turing machines \cite{deutsch 85} have
been introduced but there has been no quantum analogue to Church's $\lambda $%
-calculus. The $\lambda $-calculus has served as the basis for most
programming languages since it was introduced by Alonzo Church \cite{church}
in 1936. It and other calculi make the implicit assumption that a term may
be innocuously observed at any point. Such an assumption is hard to separate
from a system of rewriting rules because to rewrite a term, you must have
read it. However, as has been pointed out by Deutsch \cite{deutsch 92}, any
physical system is a computer. We may prepare it in some state, let it
evolve according to its dynamics, and observe it periodically. Here, the
notion of observation is crucial. One of the goals of these calculi is to
make observation explicit in the formalism itself.

The intension of the $\lambda ^{p}$- and the $\lambda ^{q}$-calculi is to
formalize computation on the level of \emph{potentia} discussed by
Heisenberg \cite{heisenberg}. Heisenberg's quantum reality is a two-world
model. One world is the world of \emph{potentia}, events that haven't
happened but could. The other world is the world of actual events that have
occured and been observed. A goal of these calculi is to allow easy
expression of algorithms that exist and operate in the world of \emph{%
potentia} yet are, at conclusion, observed.

To this end, collections (distributions and superpositions) should be
thought of with the following intuition. A collection is a bunch of terms
that do not communicate with each other. When the collection is observed, at
most one term in each collection will be the result of the observation. In
the $\lambda ^{q}$-calculus, the terms in a collection have signs, but still
do not communicate with each other. The observation process somehow removes
oppositely-signed terms. The key point is that in neither calculus can one
write a term that can determine if it is part of a collection, how big the
collection is, or even if its argument is part of a collection. Collections
can be thought of as specifications of parallel terms whose execution does
not depend on the execution of other terms in the same collection.

\section{The Lambda Calculus}

This section is a review of the $\lambda $-calculus and a reference for
later calculi.

The $\lambda $-calculus is a calculus of functions. Any computable
single-argument function can be expressed in the $\lambda $-calculus. Any
computable multiple-argument function can be expressed in terms of
computable single-argument functions. The $\lambda $-calculus is useful for
encoding functions of arbitrary arity that return at most one output for
each input. In particular, the $\lambda $-calculus can be used to express
any (computable) \emph{algorithm}. The definition of algorithm is usually
taken to be Turing-computable.

\subsection{Syntax}

The following grammar specifies the syntax of the $\lambda $-calculus. 
\begin{equation}
\begin{tabular}{|ll|}
\hline
$
\begin{array}{ll}
x & \in \text{\emph{Variable}} \\ 
M & \in \text{\emph{LambdaTerm}} \\ 
w & \in \text{\emph{Wff}}
\end{array}
$ & $
\begin{array}{l}
\text{Variables} \\ 
\text{Terms of the }\lambda \text{-calculus} \\ 
\text{Well-formed formulas of the }\lambda \text{-calculus}
\end{array}
$ \\ 
&  \\ 
$
\begin{array}{lll}
M & ::= & x \\ 
& \,\,\,| & M_{1}M_{2} \\ 
& \,\,\,| & \lambda x.M \\ 
&  &  \\ 
w & ::= & M_{1}=M_{2}
\end{array}
$ & $
\begin{array}{l}
\text{variable} \\ 
\text{application} \\ 
\text{abstraction} \\ 
\\ 
\text{well-formed formula}
\end{array}
$ \\ \hline
\end{tabular}
\label{lambda syntax}
\end{equation}

To be strict, the subscripts above should be removed (e.g., the rule for
well-formed formulas should read\ $w::=M=M$) because $M_{1}$ and $M_{2}$ are
not defined. However, we will maintain this incorrect notation to emphasize
that the terms need not be identical.

With this abuse of notation, we can easily read the preceding definition as:
a $\lambda $-term is a variable, or an application of two terms, or the
abstraction of a term by a variable. A well-formed formula of the $\lambda $%
-calculus is a $\lambda $-term followed by the equality sign followed by a
second $\lambda $-term.

We also adopt some syntactic conventions. Most importantly, parentheses
group subexpressions. Application is taken to be left associative so that
the term $MNP$ is correctly parenthesized as $\left( MN\right) P$ and not as 
$M\left( NP\right) .$ The scope of an abstraction extends as far to the
right as possible, for example up to a closing parenthesis, so that the term 
$\lambda x.xx$ is correctly parenthesized as $\left( \lambda x.xx\right) $
and not as $\left( \lambda x.x\right) x.$

\subsection{Substitution}

We will want to substitute arbitrary $\lambda $-terms for variables. We
define the substitution operator, notated$~M\left[ N/x\right] $ and read ``$%
M $ with all free occurences of $x$ replaced by $N$.'' The definition of the
free and bound variables of a term are standard. The set of free variables
of a term $M$ is written $FV\left( M\right) $. There are six rules of
substitution, which we write for reference. 
\begin{equation}
\begin{array}{ll}
1.\;x\left[ N/x\right] \equiv N &  \\ 
2.\;y\left[ N/x\right] \equiv y & \text{for variables }y\not{\equiv}x \\ 
3.\;\left( PQ\right) \left[ N/x\right] \equiv \left( P\left[ N/x\right]
\right) \left( Q\left[ N/x\right] \right) &  \\ 
4.\;\left( \lambda x.P\right) \left[ N/x\right] \equiv \lambda x.P &  \\ 
5.\;\left( \lambda y.P\right) \left[ N/x\right] \equiv \lambda y.\left(
P\left[ N/x\right] \right) & \text{if }y\not{\equiv}x\text{ and }y\notin
FV\left( N\right) \\ 
6.\;\left( \lambda y.P\right) \left[ N/x\right] \equiv \lambda z.\left(
P\left[ z/y\right] \left[ N/x\right] \right) & \text{if }y\not{\equiv}x\text{
and }y\in FV\left( N\right) \text{ } \\ 
& \text{and }z\notin FV(P)\bigcup FV\left( N\right)
\end{array}
\label{substitution}
\end{equation}

This definition will be extended in both subsequent calculi.

\subsection{Reduction}

\label{notions of reduction}The concept of \emph{reduction} seeks to
formalize rewriting rules. Given a relation $R$ between terms, we may define
the one-step reduction relation, notated$~\rightarrow _{R},$ that is the
contextual closure of $R.$ We may also define the reflexive, transitive
closure of the one-step reduction relation, which we call $R$-reduction and
notate$~\twoheadrightarrow _{R},$ and the symmetric closure of $R$%
-reduction, called $R$-interconvertibility and notated$~=_{R}.$

The essential notion of reduction for the $\lambda $-calculus is called $%
\beta $-reduction. It is based on the $\beta $-relation, which is the
formalization of function invocation. 
\begin{equation}
\beta \triangleq \left\{ \left( \left( \lambda x.M\right) N,M\left[
N/x\right] \right) \,\,\,|\,\,\,M,N\in LambdaTerm,\,x\in Variable\right\}
\label{beta}
\end{equation}

There is also the $\alpha $-relation that holds of terms that are identical
up to a consistent renaming of variables. 
\begin{equation}
\alpha \triangleq \left\{ \left( \lambda x.M,\lambda y.M\left[ y/x\right]
\right) \,\,\,|\,\,\,M\in LambdaTerm,\,y\notin FV\left( M\right) \right\}
\end{equation}
We will use this only sparingly.

\subsection{Evaluation Semantics}

By imposing an evaluation order on the reduction system, we are providing
meaning to the $\lambda $-terms. The evaluation order of a reduction system
is sometimes called an operational semantics or an evaluation semantics for
the calculus. The evaluation relation is typically denoted $\rightsquigarrow
.$

We use call-by-value evaluation semantics. A \emph{value }is the result
produced by the evaluation semantics. Call-by-value semantics means that the
body of an abstraction is not reduced but arguments are evaluated before
being passed into abstractions.

There are two rules for the call-by-value evaluation semantics of the $%
\lambda $-calculus. 
\begin{eqnarray*}
&&\frac {}{v\rightsquigarrow v}\text{(Refl)\qquad \qquad (for }v\text{ a
value)} \\
&&\frac{M\rightsquigarrow \lambda x.P\quad N\rightsquigarrow N^{\prime
}\quad P\left[ N^{\prime }/x\right] \rightsquigarrow v}{MN\rightsquigarrow v}%
\text{(Eval)}
\end{eqnarray*}

\subsection{Reference Terms}

The following $\lambda $-terms are standard and are provided as reference
for later examples.

Numbers are represented as Church numerals. 
\begin{eqnarray}
\underline{0} &\equiv &\lambda x.\lambda y.y \\
\underline{n} &\equiv &\lambda x.\lambda y.x^{n}y
\end{eqnarray}

\noindent where the notation $x^{n}y$ means $n$ right-associative
applications of $x$ onto $y.$ It is abbreviatory for the term $
\begin{array}{l}
\underbrace{x(x(\cdots (x}y))) \\ 
\,n\text{ times}
\end{array}
.$ When necessary, we can extend Church numerals to represent both positive
and negative numbers. For the remainder of the terms, we will not provide
definitions. The predecessor of Church numerals is written $\underline{\text{%
P}}.$ The successor is written $\underline{\text{S}}.$

The conditional is written $\underline{\text{IF}}.$ If its first argument is
truth, written $\underline{\text{T}},$ then it returns its second argument.
If its first argument is falsity, written $\underline{\text{F}},$ then it
returns its third argument. A typical predicate is $\underline{\text{0?}}$
which returns $\underline{\text{T}}$ if its argument is the Church numeral $%
\underline{\text{0}}$ and $\underline{\text{F}}$ if it is some other Church
numeral.

The fixed-point combinator is written $\underline{\text{Y}}.$ The primitive
recursive function-building term is written $\underline{\text{PRIM-REC}}$
and it works as follows. If the value of a function $f$ at input $n$ can be
expressed in terms of $n-1$ and $f\left( n-1\right) ,$ then that function $f$
is primitive recursive, and it can be generated by providing $\underline{%
\text{PRIM-REC}}$ with the function that takes the inputs $n-1$ and $f\left(
n-1\right) $ to produce $f\left( n\right) $ and with the value of $f$ at
input $0.$ For example, the predecessor function for Church numerals can be
represented as $\underline{\text{P}}\equiv \underline{\text{PRIM-REC}}%
\,\left( \lambda x.\lambda y.x\right) \,\underline{\text{0}}.$

\section{The Lambda-P Calculus}

The $\lambda ^{p}$-calculus is an extension of the $\lambda $-calculus that
permits the expression of \emph{randomized} algorithms. In contrast with a
computable algorithm which returns at most one output for each input, a
randomized algorithm returns a \emph{distribution} of answers from which we
sample. There are several advantages to randomized algorithms.

\begin{enumerate}
\item  Randomized algorithms can provide truly random number generators
instead of relying on pseudo-random number generators that work only because
the underlying pattern is difficult to determine.

\item  Because they can appear to generate random numbers arbitrarily,
randomized algorithms can model random processes.

\item  Given a problem of finding a suitable solution from a set of
possibilities, a randomized algorithm can exhibit the effect of choosing
random elements and testing them. Such algorithms can sometimes have an 
\emph{expected }running time which is considerably shorter than the running
time of the computable algorithm that tries every possibility until it finds
a solution.
\end{enumerate}

\subsection{Syntax}

\label{section:lambda-p syntax}The following grammar describes the $\lambda
^{p}$-calculus. 
\begin{equation}
\begin{tabular}{|ll|}
\hline
$
\begin{array}{ll}
x & \in \text{\emph{Variable}} \\ 
M & \in \text{\emph{LambdaPTerm}} \\ 
w & \in \text{\emph{WffP}}
\end{array}
$ & $
\begin{array}{l}
\text{Variables} \\ 
\text{Terms of the }\lambda ^{p}\text{-calculus} \\ 
\text{Well-formed formulas of the }\lambda ^{p}\text{-calculus}
\end{array}
$ \\ 
&  \\ 
$
\begin{array}{lll}
M & ::= & x \\ 
& \,\,\,| & M_{1}M_{2} \\ 
& \,\,\,| & \lambda x.M \\ 
& \,\,\,| & M_{1},M_{2} \\ 
&  &  \\ 
w & ::= & M_{1}=M_{2}
\end{array}
$ & $
\begin{array}{l}
\text{variable} \\ 
\text{application} \\ 
\text{abstraction} \\ 
\text{collection} \\ 
\\ 
\text{well-formed formula}
\end{array}
$ \\ \hline
\end{tabular}
\newline
\label{lambda-p syntax}
\end{equation}

Note that this grammar differs from the $\lambda $-calculus only in the
addition of the fourth rule for terms. Therefore, all $\lambda $-terms can
be viewed as $\lambda ^{p}$-terms.

A $\lambda ^{p}$-term is a variable, or an application of two terms, or the
abstraction of a term by a variable, or a collection of two terms. It
follows that a term may be a collection of a term and another collection, so
that a term may actually have many nested collections.

We adhere to the same parenthesization and precedence rules as the $\lambda $%
-calculus with the following addition:\ collection is of lowest precedence
and the comma is right associative. This means that the expression $\lambda
x.x,z,y$ is correctly parenthesized as $\left( \lambda x.x\right) ,(z,y)$.

We introduce abbreviatory notation for collections. Let us write $\left[
M_{i}^{i\in S}\right] $ for the collection of terms $M_{i}$ for all $i$ in
the finite, ordered set $S$ of natural numbers. We will write $a..b$ for the
ordered set $\left( a,a+1,\ldots ,b\right) .$ In particular, $\left[
M_{i}^{i\in 1..n}\right] $ represents $M_{1},M_{2},\ldots ,M_{n}$ and $%
\left[ M_{i}^{i\in n..1}\right] $ represents $M_{n},M_{n-1},\ldots ,M_{1}$.
More generally, let us allow multiple iterators in arbitrary contexts. Then,
for instance, 
\[
\left[ \lambda x.M_{i}^{i\in 1..n}\right] \equiv \lambda x.M_{1},\lambda
x.M_{2},\ldots ,\lambda x.M_{n} 
\]
and 
\[
\left[ M_{i}^{i\in 1..m}N_{j}^{j\in 1..n}\right] \equiv 
\begin{array}{c}
M_{1}N_{1},M_{1}N_{2},\ldots ,M_{1}N_{n}, \\ 
M_{2}N_{1},M_{2}N_{2},\ldots ,M_{2}N_{n}, \\ 
\vdots \\ 
M_{m}N_{1},M_{m}N_{2},\ldots ,M_{m}N_{n}
\end{array}
. 
\]

Note that $\left[ \lambda x.M_{i}^{i\in 1..n}\right] $ and $\lambda x.\left[
M_{i}^{i\in 1..n}\right] $ are not the same term. The former is a collection
of abstractions while the latter is an abstraction with a collection in its
body. Finally, we allow this notation to hold of non-collection terms as
well by identifying $\left[ M_{i}^{i\in 1..1}\right] $ with $M_{1}$ even if $%
M_{1}$ is not a collection. \noindent To avoid confusion, it is important to
understand that although this ``collection'' notation can be used for
non-collections, we do not extend the definition of the word \emph{%
collection. }A \emph{collection} is still the syntactic structure defined in
grammar (\ref{lambda-p syntax}).

With these additions, every term can be written in this bracket form. In
particular, we can write a collection as $\left[ \left[ M_{i}^{i\in
S_{i}}\right] _{j}^{j\in S}\right] ,$ or a collection of collections.
Unfortunately, collections can be written in a variety of ways with this
notation. The term $M,N,P$ can be written as $\left[ M_{i}^{i\in
1..3}\right] $ if $M_{1}\equiv M$ and $M_{2}\equiv N$ and $M_{3}\equiv P;$
as $\left[ M_{i}^{i\in 1..2}\right] $ if $M_{1}\equiv M$ and $M_{2}\equiv
N,P;$ or as $\left[ M_{i}^{i\in 1..1}\right] $ if $M_{1}\equiv M,N,P.$
However, it cannot be written as $\left[ M_{i}^{i\in 1..4}\right] $ for any
identification of the $M_{i}.$ This observation inspires the following
definition.

\begin{definition}
\label{dfn: cardinality}The \emph{cardinality} of a term $M,$ notated$%
~\left| M\right| ,$ is that number $k$ for which $\left[ M_{i}^{i\in
1..k}\right] \equiv M$ for some identification of the $M_{i}$ but $\left[
M_{i}^{i\in 1..\left( k+1\right) }\right] \not{\equiv}M$ for any
identification of the $M_{i}$.
\end{definition}

\noindent Note that the cardinality of a term is always strictly positive.

\subsection{Alternative Syntax}

\label{alternative syntax}We present an alternative syntax for the $\lambda
^{p}$-calculus that is provably equivalent to the one given above under
certain assumptions. We will call the temporary calculus whose syntax we
define below the $\lambda ^{p^{\prime }}$-calculus to distinguish it from
the one we will ultimately adopt.

The following grammar describes the syntax of the $\lambda ^{p^{\prime }}$%
-calculus. 
\begin{equation}
\begin{tabular}{|ll|}
\hline
$
\begin{array}{ll}
x & \in \text{\emph{Variable}} \\ 
M & \in \text{\emph{LambdaTerm}} \\ 
C & \in \text{\emph{LambdaP}}^{\prime }\text{\emph{Term}} \\ 
w & \in \text{\emph{WffP}}^{\prime }
\end{array}
$ & $
\begin{array}{l}
\text{Variables} \\ 
\text{Terms of the }\lambda \text{-calculus} \\ 
\text{Terms of the }\lambda ^{p^{\prime }}\text{-calculus} \\ 
\text{Well-formed formulas of the }\lambda ^{p^{\prime }}\text{-calculus}
\end{array}
$ \\ 
&  \\ 
$
\begin{array}{lll}
M & ::= & x \\ 
& \,\,\,| & M_{1}M_{2} \\ 
& \,\,\,| & \lambda x.M \\ 
&  &  \\ 
C & ::= & M \\ 
& \,\,\,| & C_{1},C_{2} \\ 
& \,\,\,| & C_{1}C_{2} \\ 
&  &  \\ 
w & ::= & C_{1}=C_{2}
\end{array}
$ & $
\begin{array}{l}
\text{variable} \\ 
\text{application} \\ 
\text{abstraction} \\ 
\\ 
\text{term} \\ 
\text{construction} \\ 
\text{collection application} \\ 
\\ 
\text{well-formed formula}
\end{array}
$ \\ \hline
\end{tabular}
\newline
\label{lambda-p' syntax}
\end{equation}

The syntax of the $\lambda ^{p}$-calculus, grammar (\ref{lambda-p' syntax}),
allows the same terms and well-formed formulas as the syntax of the $\lambda
^{p^{\prime }}$-calculus, grammar (\ref{lambda-p syntax}), if we identify
abstractions of collections with the appropriate collection of abstractions,
and applications of collections with collections of applications.

\begin{theorem}
If we have, in the $\lambda ^{p}$-calculus, that 
\begin{equation}
\left[ \lambda x.M_{i}^{i\in 1..n}\right] \equiv \lambda x.\left[
M_{i}^{i\in 1..n}\right]  \label{abstraction identity}
\end{equation}
and 
\begin{equation}
\left[ M_{i}^{i\in 1..m}N_{j}^{j\in 1..n}\right] \equiv \left[ M_{i}^{i\in
1..m}\right] \left[ N_{j}^{j\in 1..n}\right] \noindent
\label{application identity}
\end{equation}
then $w$ is a well-formed formula in the $\lambda ^{p}$-calculus if and only
if it is a well-formed formula in the $\lambda ^{p^{\prime }}$-calculus.
\end{theorem}

\proof%
It is sufficient to show that an arbitrary $\lambda ^{p^{\prime }}$-term is
a $\lambda ^{p}$-term and vice versa.

First we show that an arbitrary $\lambda ^{p^{\prime }}$-term is a $\lambda
^{p}$-term by structural induction. From (\ref{lambda-p' syntax}), a $%
\lambda ^{p^{\prime }}$-term $C$ is either a $\lambda $-term, a
construction, or a collection application. If $C$ is a $\lambda $-term, then
it is a $\lambda ^{p}$-term. If it is a construction $C_{1},C_{2}$, then $C$
is a $\lambda ^{p}$-collection term $C_{1},C_{2}$ because $C_{1}$ and $C_{2}$
are $\lambda ^{p}$-terms by the induction hypothesis. Finally, if $C$ is a
collection application $C_{1}C_{2}$, then, since $C_{1}$ and $C_{2}$ are $%
\lambda ^{p}$-terms by the induction hypothesis and $C_{1}C_{2}$ is a $%
\lambda ^{p}$-application term, $C$ is a $\lambda ^{p}$-term.

Now we show that an arbitrary $\lambda ^{p}$-term is a $\lambda ^{p^{\prime
}}$-term by structural induction. If $M$ is a $\lambda ^{p}$-term, it is
either a variable, an application, a collection, or an abstraction. If $M$
is a variable, then it is a $\lambda $-term and therefore a $\lambda
^{p^{\prime }}$-term. If $M$ is an application $PQ,$ then by the induction
hypothesis $P$ and $Q$ are $\lambda ^{p^{\prime }}$-terms, so that $PQ$ is a 
$\lambda ^{p^{\prime }}$-collection application and $M$ is a $\lambda
^{p^{\prime }}$-term. If $M$ is a collection, then by the induction
hypothesis it is a collection of $\lambda ^{p}$-terms that are $\lambda
^{p^{\prime }}$-terms, so that $M$ is also a $\lambda ^{p^{\prime }}$%
-construction.

If $M$ is an abstraction $\lambda x.N,$ then by the induction hypothesis, $N$
is a $\lambda ^{p^{\prime }}$-term. Therefore, $N$ is either a $\lambda $%
-term, a construction, or a collection application. If $N$ is a $\lambda $%
-term, then $M$ is a $\lambda ^{p^{\prime }}$-term. If $N$ is a $\lambda
^{p^{\prime }}$-construction, then it is also a $\lambda ^{p}$-collection
term by the first part of this proof. Therefore, $M$ is an abstraction over
a collection, and by assumption (\ref{abstraction identity})\ is identical
to a collection over abstractions. By the induction hypothesis, each of the
abstractions in the collection is a $\lambda ^{p^{\prime }}$-term, so the
collection itself is a $\lambda ^{p^{\prime }}$-construction. Therefore, $M$
is a $\lambda ^{p^{\prime }}$-term. Finally, if $N$ is a collection
application, then by assumption (\ref{application identity}) it is identical
to a collection of applications. Therefore, $N$ is a $\lambda ^{p}$%
-collection term. By the same reasoning as in the previous case, it follows
that $M$ is a $\lambda ^{p^{\prime }}$-term.

This completes the proof.%
\endproof%

The $\lambda ^{p}$-calculus seems more expressive than the $\lambda
^{p^{\prime }}$-calculus because it allows terms to be collections of other
terms. We have seen that with the two assumptions (\ref{abstraction identity}%
)\ and (\ref{application identity}), the two calculi are equally expressive.
Without these assumptions, some abstractions can be expressed in the $%
\lambda ^{p}$-calculus that cannot be expressed in the $\lambda ^{p^{\prime
}}$-calculus. Are these assumptions justifiable?

The first assumption (\ref{abstraction identity}) states that the
abstraction of a collection is syntactically identical to the collection of
the abstractions. For example, the term $\lambda x.\left( x,xx\right) $ is
claimed to be identical to the term $\lambda x.x,\lambda x.xx.$ Given our
intuitive understanding of what these terms represent, it is indisputable
that these two terms are equal in a semantic sense. Applying each term to
arbitrary inputs ought to yield statistically indistinguishable results.
However, the question is:\ should we identify these terms on a syntactic
level? Certainly the two $\lambda $-terms $\lambda x.x$ and $\lambda
x.\left( \lambda y.y\right) x$ are semantically equivalent, but we do not
identify them on a syntactic level.

The other assumption (\ref{application identity}) states that the
application of two collections is the collection of all possible
applications of terms in the two collections. For example, the term $\left(
M,N\right) (P,Q)$ is claimed to be identical to the term $MP,MQ,NP,NQ.$
Again, these two would, given our intuition, be equal in the statistical
sense, but should they represent the same syntactic structure? Even if the
two terms represent the same thing in the real world, that is, if they share
the same denotation, it does not follow that they should be syntactically
identical. $\beta $-reduction preserves denotation, but we do not say that a
term and what it reduces to are syntactically identical.

On the other hand, we do want to identify some terms that are written
differently. For example, the order of terms in a collection ought not
distinguish terms. The terms $M,N$ and $N,M$ should be identified, given our
intuitive understanding of what these terms mean. Identifying unordered
terms is not uncommon and is done in other calculi \cite{abadi-cardelli}.
How do we decide whether to identify certain pairs of terms or to define a
notion of reduction for them?

We want to identify terms when the differences do not affect computation and
result from the limiting nature of writing. Identifying collections with
different orders is a workaround for the sequentiality and specificity of
the comma operator. Together with the grammar, such an identification
clarifies the terms of discourse. However, this reasoning does not apply to
the assumptions (\ref{abstraction identity}) and (\ref{application identity}%
) because these assumptions are trying to identify terms that bear only a
semantic relationship to each other.

Much as we refrain from identifying a term with its $\beta $-reduced form,
we do not want to identify an application (abstraction)\ of collections with
a collection of applications (abstractions). We may choose instead to
capture this relationship in the form of a relation and associated
reductions. Such is the approach we will adopt here.

Of the two grammars, we choose the $\lambda ^{p}$-calculus because it is the
more general one. We will define the $\gamma $-relation to hold of an
application of collections and a collection of applications. However, we
will neither identify nor provide a relation for the analogous abstraction
relationship of the pair of terms identified in assumption (\ref{abstraction
identity}), because such a step would be redudant. By the observation
function we will define in \S \ref{lambda-p observation}, observing an
abstraction of collections is tantamount to observing a collection of
abstractions, so no new power or expressibility would be gained.

\subsection{Syntactic Identities}

We define substitution of terms in the $\lambda ^{p}$-calculus as an
extension of substitution of terms in the $\lambda $-calculus. In addition
to the six rules of the $\lambda $-calculus, we introduce one for
collections. 
\begin{equation}
\left( P,Q\right) \left[ N/x\right] \equiv \left( P\left[ N/x\right]
,Q\left[ N/x\right] \right)  \label{substitution-p}
\end{equation}

We identify terms that are collections but with a possibly different
ordering. We also identify nested collections with the top-level collection.
The motivation for this is the conception that a collection is an unordered
set of terms. Therefore we will not draw a distinction between a set of
terms and a set of a set of terms.

We adopt the following axiomatic judgement rules. 
\begin{eqnarray*}
&&\dfrac {}{M,N\equiv N,M}\text{(ClnOrd)} \\
&&\dfrac {}{\left( M,N\right) ,P\equiv M,(N,P)}\text{(ClnNest)}
\end{eqnarray*}

With these axioms, ordering and nesting become innocuous. As an example here
is the proof that $A,(B,C),D\equiv A,C,B,D.$ For clarity, we parenthesize
fully and underline the affected term in each step. 
\[
\begin{array}{llll}
\underline{A,((B,C),D)} & \equiv & ((\underline{B,C}),D),A & \text{(ClnOrd)}
\\ 
& \equiv & (\underline{(C,B),D}),A & \text{(ClnOrd)} \\ 
& \equiv & \underline{(C,(B,D)),A} & \text{(ClnNest)} \\ 
& \equiv & A,(C,(B,D)) & \text{(ClnOrd)}
\end{array}
\]

We now show that ordering and parenthesization are irrelevant in general.

\begin{theorem}
\label{theorem:order/paren invariance}If the $n$ ordered sets $S_{i},$ $%
1\leq i\leq n,$ are distinct and $\Pi $ is a permutation of the ordered set $%
1..n$, then $\left[ \left[ M_{i}^{i\in S_{i}}\right] _{j}^{j\in 1..n}\right]
\equiv \left[ M_{i}^{i\in S_{1}S_{2}\cdots S_{n}}\right] $ and $\left[
M_{i}^{i\in 1..n}\right] \equiv \left[ M_{i}^{i\in \Pi }\right] ,$ where
juxtaposition of ordered sets denotes extension (e.g., $\left( 1..3\right)
\left( 5..7\right) =\left( 1,2,3,5,6,7\right) $).
\end{theorem}

\proof%
We prove this theorem by induction on $n.$

There are two base cases. When $n=1,$ the claim holds trivially. When $n=2,$
the claim follows from the $\left( \text{ClnOrd}\right) $ axiom.

For the inductive case, we consider $M\equiv \left[ M_{i}^{i\in 1..\left(
n+1\right) }\right] \equiv M_{1},\left[ M_{i}^{i\in 2..\left( n+1\right)
}\right] $ and assume the claim holds for all collections of $n$ or fewer
terms, and that $n\geq 2.$

To show parenthesization invariance, we write $M\equiv P,Q$ where $P\equiv
\left[ P_{i}^{i\in 1..n}\right] $ and $Q$ are collections. It will be
sufficient to show that $M\equiv \left[ P_{i}^{i\in 1..(n-1)}\right] ,\left(
P_{n},Q\right) $. By the induction hypothesis, we may parenthesize $P$
arbitrarily. We choose to parenthesize $P$ left-associatively as 
\[
\left( \left( \left( P_{1},P_{2}\right) ,P_{3}\,\cdots \,P_{n-2}\right)
,P_{n-1}\right) ,P_{n}. 
\]
Then, by the $\left( \text{ClnNest}\right) $ axiom, $M\equiv \left( \left(
\left( P_{1},P_{2}\right) ,P_{3}\,\cdots \,P_{n-2}\right) ,P_{n-1}\right)
,\left( P_{n},Q\right) ,$which is identical to $\left[ P_{i}^{i\in
1..(n-1)}\right] ,\left( P_{n},Q\right) $ by the reordering allowed by the
induction hypothesis. This completes the proof of parenthesization
invariance.

To show reordering invariance, note that the permutation $\Pi $ of $%
1..\left( n+1\right) $ either has $1$ as its first element or it does not.
If it does, then by the induction hypothesis, $\left[ M_{i}^{i\in 2..\left(
n+1\right) }\right] $ can be reordered in an arbitrary manner, so that $%
M\equiv \left[ M_{i}^{i\in \Pi }\right] .$ If it does not, then the first
element of $\Pi $ is an integer $k$ between $2$ and $n-1.$ By the induction
hypothesis, we can reorder $\left[ M_{i}^{i\in 2..\left( n+1\right) }\right] 
$ as $M_{k},\left[ M_{i}^{i\in 2..\left( n+1\right) -\{k]}\right] ,$ where
we write $2..\left( n+1\right) -k$ for the ordered set $\left( 2,3,\ldots
,k-1,k+1,\ldots ,n+1\right) .$ Then, underlying the affected term, we get 
\[
\begin{array}{lll}
M & \equiv \underline{M_{1},M_{k},\left[ M_{i}^{i\in 2..\left( n+1\right)
-\{k]}\right] } & \text{by reordering} \\ 
& \equiv \left( \underline{M_{1},M_{k}}\right) ,\left[ M_{i}^{i\in 2..\left(
n+1\right) -\{k]}\right] & \text{by the }\left( \text{ClnNest}\right) \text{
axiom} \\ 
& \equiv \underline{\left( M_{k},M_{1}\right) ,\left[ M_{i}^{i\in 2..\left(
n+1\right) -\{k]}\right] } & \text{by the }\left( \text{ClnOrd}\right) \text{
axiom} \\ 
& \equiv M_{k},M_{1},\left[ M_{i}^{i\in 2..\left( n+1\right) -\{k]}\right] & 
\text{by the }\left( \text{ClnNest}\right) \text{ axiom}
\end{array}
\]

\noindent Then, by the induction hypothesis, $M_{1},\left[ M_{i}^{i\in
2..\left( n+1\right) -\{k]}\right] $ can be reordered arbitrarily so that $M$
can be reordered to fit the permutation $\Pi $. This completes the proof of
reordering invariance and of the theorem.%
\endproof%

Aside, it no longer matters that we took the comma to be right associative
since, with these rules, any arbitrary parenthesization of a collection does
not change the syntactic structure.

Because of this theorem, we can alter the abbreviatory notation and allow
arbitrary unordered sets in the exponent. This allows us to write, for
instance, $\left[ M_{i}^{i\in 1..n-\{j\}}\right] \equiv M_{1},M_{2},\ldots
,M_{j-1},M_{j+1},\ldots ,M_{n}$ where $a..b$ is henceforth taken to be the
unordered set $\left\{ a,a+1,\ldots ,b\right\} $ and the subtraction in the
exponent represents set difference.

This also subtly alters the definition of \emph{cardinality }(\ref{dfn:
cardinality}). Whereas before the cardinality of a term like $\left(
x,y\right) ,z$ was 2, because of this theorem, it is now 3. Because every $%
\lambda ^{p}$-term is finite, the cardinality is well-defined.

\subsection{Reductions}

The relation of collection application is called the $\gamma $-relation. It
holds of a term that is an application at least one of whose operator or
operand is a collection, and the term that is the collection of all possible
pairs of applications.

\begin{equation}
\gamma ^{p}\triangleq \left\{ 
\begin{array}{l}
\left( \left[ M_{i}^{i\in 1..m}\right] \left[ N_{j}^{j\in 1..n}\right]
,\left[ M_{i}^{i\in 1..m}N_{j}^{j\in 1..n}\right] \right) \\ 
\text{such that }M_{i},N_{j}\in LambdaPTerm,\,m>1\text{ or }n>1
\end{array}
\right\}  \label{gamma-p}
\end{equation}

\noindent The $\gamma $-relation is our solution to the concerns of \S \ref
{alternative syntax} regarding claim (\ref{application identity}). We will
omit the superscript except to disambiguate from the $\gamma $-relation of
the $\lambda ^{q}$-calculus.

We generate the reduction relations as described in \S \ref{notions of
reduction} to get the relations of $\gamma $ -reduction in one step $%
\rightarrow _{\gamma },$ $\gamma $-reduction $\twoheadrightarrow _{\gamma },$
and $\gamma $-interconvertibility $=_{\gamma }.$

The $\gamma $-relation is Church-Rosser.

\begin{theorem}
\label{thm: gamma-p CR}For $\lambda ^{p}$-terms $M,R,S,$ if $%
M\twoheadrightarrow _{\gamma }R$ and $M\twoheadrightarrow _{\gamma }S$ then
there exists a $\lambda ^{p}$-term $T$ such that $R\twoheadrightarrow
_{\gamma }T$ and $S\twoheadrightarrow _{\gamma }T.$
\end{theorem}

\noindent This is shown in the standard way by proving the associated strip
lemma.

As a result of this theorem, $\gamma $-normal forms, when they exist, are
unique. We now show that all terms have $\gamma $-normal forms.

\begin{theorem}
\label{thm: gamma-p NF exists}For every $\lambda ^{p}$-term $M$ there exists
another $\lambda ^{p}$-term $N$ such that $M\twoheadrightarrow _{\gamma }N$
and $N$ has no $\gamma $-redexes.
\end{theorem}

\proof%
The proof is by structural induction on $M$.

If $M\equiv x$ is a variable, there are no $\gamma $-redexes, so $N\equiv M$.

If $M\equiv \lambda x.P$ is an abstraction, then the only $\gamma $-redexes,
if any, are in $P.$ By the induction hypothesis, there exists a term $%
P^{\prime }$ such that $P\twoheadrightarrow _{\gamma }P^{\prime }$ and $%
P^{\prime }$ has no $\gamma $-redexes. Then $M\twoheadrightarrow _{\gamma
}\lambda x.P^{\prime }\equiv N$ and $N$ has no $\gamma $-redexes either.

If $M\equiv PQ$ is an application, then there exist terms $P^{\prime
},Q^{\prime }$ such that $P\twoheadrightarrow _{\gamma }P^{\prime }$ and $%
Q\twoheadrightarrow _{\gamma }Q^{\prime }$ and neither $P^{\prime }$ nor $%
Q^{\prime }$ have $\gamma $-redexes. Let $\left[ P_{i}^{i\in 1..\left|
P^{\prime }\right| }\right] \equiv P^{\prime }$ and $\left[ Q_{i}^{i\in
1..\left| Q^{\prime }\right| }\right] \equiv Q^{\prime }.$ Then $%
M\twoheadrightarrow _{\gamma }P^{\prime }Q^{\prime }\rightarrow _{\gamma
}\left[ P_{i}^{i\in 1..\left| P^{\prime }\right| }Q_{j}^{j\in 1..\left|
Q^{\prime }\right| }\right] \equiv N$ where none of the $P_{i}$ or $Q_{j}$
are collections, by the definition of cardinality. Also, since neither $%
P^{\prime }$ nor $Q^{\prime }$ had $\gamma $-redexes, none of the $P_{i}$ or 
$Q_{j}$ have $\gamma $-redexes either. Therefore, $N$ does not have any $%
\gamma $-redexes.

If $M\equiv \left[ M_{i}^{i\in 1..\left| M\right| }\right] $ is a
collection, then for each $M_{i}$ there exists a term $N_{i}$ such that $%
M_{i}\twoheadrightarrow _{\gamma }N_{i}$ and $N_{i}$ has no $\gamma $%
-redexes. Then $M\twoheadrightarrow _{\gamma }\left[ N_{i}^{i\in 1..\left|
M\right| }\right] \equiv N$ and since none of the $N_{i}$ have $\gamma $%
-redexes, neither does $N.$

This exhausts the cases and completes the proof.%
\endproof%

\noindent Therefore, all $\lambda ^{p}$-terms have normal forms and they are
unique , so we may write $\gamma \left( M\right) $ for the $\gamma $-normal
form of a $\lambda ^{p}$-term $M.$

We extend the $\beta $-relation to apply to collections. 
\begin{equation}
\beta ^{p}\triangleq \left\{ 
\begin{array}{l}
\left( \left( \lambda x.M\right) \left[ N_{i}^{i\in S}\right] ,\left[
M\left[ N_{i}^{i\in S}/x\right] \right] \right) \\ 
\text{such that }M\text{ and }\left[ N_{i}^{i\in S}\right] \in
LambdaPTerm,\,x\in Variable
\end{array}
\right\}  \label{beta-p}
\end{equation}
where $\left[ M\left[ N_{i}^{i\in S}/x\right] \right] $ is the collection of
terms $M$ with $N_{i}$ substituted for free occurrences of $x$ in $M,$ for $%
i\in S.$

One-step $\beta $-reduction in the $\lambda ^{p}$-calculus $\rightarrow
_{\beta ^{p}}$ is different from that of the $\lambda $-calculus because the
grammar is extended. (Note that we omit the superscript on $\beta $%
-reduction when there is no ambiguity about which calculus is under
consideration.) Therefore, we need to prove that the $\beta $-relation is
still Church-Rosser in the $\lambda ^{p}$-calculus. This is easy to do but
not helpful because the appropriate notion of reduction in the $\lambda ^{p}$%
-calculus is not $\beta $-reduction, but $\beta $-reduction with $\gamma $%
-reduction to normal form after each $\beta $-reduction step.

We define the relation $\beta \gamma $ which is just like the $\beta $%
-relation except that the resultant term is in $\gamma $-normal form. 
\begin{equation}
\beta \gamma \triangleq \left\{ \left( M,\gamma \left( N\right) \right)
\,\,\,|\,\,\,\left( M,N\right) \in \beta ^{p}\right\}  \label{beta-gamma-p}
\end{equation}

The $\beta \gamma $-relation is Church-Rosser.

\begin{theorem}
\label{thm: beta-gamma-p CR}For $\lambda ^{p}$-terms $M,R,S,$ if $%
M\twoheadrightarrow _{\beta }R$ and $M\twoheadrightarrow _{\beta }S$ then
there exists a $\lambda ^{p}$-term $T$ such that $R\twoheadrightarrow
_{\beta }T$ and $S\twoheadrightarrow _{\beta }T.$
\end{theorem}

\noindent Again the proof follows the standard framework.

\subsection{Evaluation Semantics}

We extend the call-by-value evaluation semantics of the $\lambda $-calculus.

There are three rules for the call-by-value evaluation semantics of the $%
\lambda ^{p}$-calculus. We modify the definition of a value $v$ to enforce
that $v$ has no $\gamma $-redexes.

\begin{eqnarray*}
&&\dfrac {}{v\rightsquigarrow v}\text{(Refl)\qquad \qquad (for }v\text{ a
value)} \\
&&\dfrac{\gamma \left( M\right) \rightsquigarrow \lambda x.P\quad \gamma
\left( N\right) \rightsquigarrow N^{\prime }\quad \gamma \left( P\left[
N^{\prime }/x\right] \right) \rightsquigarrow v}{MN\rightsquigarrow v}\text{%
(Eval)} \\
&&\dfrac{\gamma \left( M\right) \rightsquigarrow v_{1}\quad \gamma \left(
N\right) \rightsquigarrow v_{2}}{\left( M,N\right) \rightsquigarrow \left(
v_{1},v_{2}\right) }\text{(Coll)}
\end{eqnarray*}

\subsection{Observation}

\label{lambda-p observation}We define an observation function $\Theta $ from 
$\lambda ^{p}$-terms to $\lambda $-terms. We employ the random number
generator $RAND$, which samples one number from a given set of numbers. 
\begin{eqnarray}
\Theta \left( x\right) &=&x \\
\Theta \left( \lambda x.M\right) &=&\lambda x.\Theta \left( M\right) \\
\Theta \left( M_{1}M_{2}\right) &=&\Theta \left( M_{1}\right) \Theta \left(
M_{2}\right) \\
\Theta \left( M\equiv \left[ M_{i}^{i\in 1..\left| M\right| }\right] \right)
&=&M_{RAND(1..\left| M\right| )}
\end{eqnarray}

\noindent The function $\Theta $ is total because every $\lambda ^{p}$-term
is mapped to a $\lambda $-term. Note that for an arbitrary term $T$ we may
write $\Theta \left( T\right) =T_{RAND(S)}$ for some possibly singleton set
of natural numbers $S$ and some collection of terms $\left[ T_{i}^{i\in
S}\right] .$

\begin{definition}
\label{defn: statistically indistinguishable}We say that $\Theta \left(
M\right) =M_{RAND\left( S\right) }$ is \emph{statistically indistinguishable 
}from $\Theta \left( N\right) =N_{RAND\left( S^{\prime }\right) }$, written $%
\Theta \left( M\right) \stackrel{d}{=}\Theta \left( N\right) ,$ if there
exist positive integers $m$ and $n$ and a total, surjective mapping $\varphi 
$ between $S$ and $S^{\prime }$ such that, for each $k\in S,$ $M_{k}\equiv
N_{\varphi \left( k\right) }$ and 
\begin{equation}
\frac{\left| \left[ M_{i}^{i\in \left\{ j\,\,\,|\,\,\,M_{j}\equiv
M_{k}\right\} }\right] \right| }{\left| S\right| }=\frac{\left| \left[
N_{i}^{i\in \left\{ \varphi \left( j\right) \,\,\,|\,\,\,M_{j}\equiv
M_{k}\right\} }\right] \right| }{\left| S^{\prime }\right| }
\end{equation}
that is, if the proportion of terms in $M$ identical to $M_{k}$ is the same
as the proportion of terms in $N$ identical to $N_{\varphi \left( k\right)
}, $ for all $k\in S.$ In particular, if the mapping $\varphi $ is an
isomorphism, $\Theta \left( M\right) $ is said to be \emph{statistically
identical }to $\Theta \left( N\right) $, written $\Theta \left( M\right) 
\stackrel{D}{=}\Theta \left( N\right) .$
\end{definition}

Because the mapping $\varphi $ is total and surjective, statistical
indistinguishability is a symmetric property. Note that statististical
identity is a restatement of the theorem of parenthesization and ordering
invariance (\ref{theorem:order/paren invariance}). We now show that
observing a $\lambda ^{p}$-term is statistically indistinguishable from
observing its $\gamma $-normal form.

\begin{theorem}
\label{thm: gamma-p preserves stats. ind.}If $M\twoheadrightarrow _{\gamma
}N $ then $\Theta \left( M\right) \stackrel{d}{=}\Theta \left( N\right) .$
\end{theorem}

\proof%
The proof is by structural induction on $M$. The induction hypothesis is
stronger than required and states that if $M\twoheadrightarrow _{\gamma }N$
then $\Theta \left( M\right) $ and $\Theta \left( N\right) $ are
statistically identical.

If $M$ is a variable then $M\equiv N$ and $\Theta \left( M\right) =\Theta
\left( N\right) .$ In particular, $\Theta \left( M\right) \stackrel{D}{=}%
\Theta \left( N\right) .$

If $M\equiv \lambda x.P$ is an abstraction then $N\equiv \lambda x.P^{\prime
}$ where $P\twoheadrightarrow _{\gamma }P^{\prime }$. By the induction
hypothesis, $\Theta \left( P\right) \stackrel{D}{=}\Theta \left( P^{\prime
}\right) ,$ and by definition $\Theta \left( M\right) =\lambda x.\Theta
\left( P\right) \stackrel{D}{=}\lambda x.\Theta \left( P^{\prime }\right)
=\Theta \left( \lambda x.P^{\prime }\right) =\Theta \left( N\right) $.

If $M\equiv PQ$ is an application, then either $N$ is an application or a
collection. If $N$ is an application, then $N\equiv P^{\prime }Q^{\prime }$
and, since $P\twoheadrightarrow _{\gamma }P^{\prime }$ and $%
Q\twoheadrightarrow _{\gamma }Q^{\prime },$ we have by the induction
hypothesis that $\Theta \left( P\right) \stackrel{D}{=}\Theta \left(
P^{\prime }\right) $ with isomorphism $\varphi _{P}$ and $\Theta \left(
Q\right) \stackrel{D}{=}\Theta \left( Q^{\prime }\right) $ with isomorphism $%
\varphi _{Q}$. We write 
\begin{eqnarray*}
\Theta \left( P\right) &=&P_{RAND\left( S_{P}\right) } \\
\Theta \left( P^{\prime }\right) &=&P_{RAND\left( S_{P^{\prime }}\right)
}^{\prime } \\
\Theta \left( Q\right) &=&Q_{RAND\left( S_{Q}\right) } \\
\Theta \left( Q^{\prime }\right) &=&Q_{RAND\left( S_{Q^{\prime }}\right)
}^{\prime } \\
\Theta \left( M\right) &=&M_{RAND\left( S_{M}\right) } \\
\Theta \left( N\right) &=&N_{RAND\left( S_{N}\right) }
\end{eqnarray*}

\noindent and note that it is sufficient to exhibit an isomorphism between $%
\Theta \left( M\right) =\Theta \left( PQ\right) $ and $\Theta \left(
N\right) =\Theta \left( P^{\prime }Q^{\prime }\right) .$ Without loss of
generality, let $S_{p}=S_{p^{\prime }}=1..p$ and $S_{q}=S_{q^{\prime }}=1..q$
so we can write, for $i\in 1..\left( pq\right) ,$%
\begin{eqnarray}
M_{i} &\equiv &P_{1+\left( i-1\right) \backslash q}Q_{1+\left( i-1\right) 
\func{mod}q} \\
N_{i} &\equiv &P_{1+\left( i-1\right) \backslash q}^{\prime }Q_{1+\left(
i-1\right) \func{mod}q}^{\prime }  \label{Ni}
\end{eqnarray}
\noindent where $x\backslash y=\left\lfloor \frac{x}{y}\right\rfloor .$
Remember that by theorem (\ref{theorem:order/paren invariance}), order and
parenthesization does not matter. Given isomorphisms $\varphi _{P}$ and $%
\varphi _{Q},$ we need to find an isomorphism, $\varphi ,$ between the $%
M_{i} $ and $N_{i}.$ We rewrite 
\begin{eqnarray}
M_{i} &\equiv &P_{\varphi _{P}\left( 1+\left( i-1\right) \backslash q\right)
}^{\prime }Q_{1+\left( i-1\right) \func{mod}q} \\
&\equiv &P_{\varphi _{P}\left( 1+\left( i-1\right) \backslash q\right)
}^{\prime }Q_{\varphi _{Q}\left( 1+\left( i-1\right) \func{mod}q\right)
}^{\prime } \\
&\equiv &N_{1+\left( \varphi _{P}\left( 1+\left( i-1\right) \backslash
q\right) -1\right) q+\left( \varphi _{Q}\left( 1+\left( i-1\right) \func{mod}%
q\right) -1\right) }
\end{eqnarray}

\noindent where the last identity follows from rewriting identity (\ref{Ni})
as 
\begin{equation}
N_{1+\left( j-1\right) q+\left( k-1\right) }\equiv P_{j}Q_{k}
\end{equation}

\noindent where $j\in 1..p,\,k\in 1..q.$ Thus, the isomorphism $\varphi
\left( i\right) =1+\left( \varphi _{P}\left( 1+\left( i-1\right) \backslash
q\right) -1\right) q+\left( \varphi _{Q}\left( 1+\left( i-1\right) \func{mod}%
q\right) -1\right) $ satisfies the definition of statistical identity (\ref
{defn: statistically indistinguishable}) so $\Theta \left( M\right) 
\stackrel{D}{=}\Theta \left( N\right) .$

Finally, if $M\equiv \left[ M_{i}^{i\in 1..\left| M\right| }\right] $ is a
collection, then $N\equiv \left[ N_{i}^{i\in 1..\left| M\right| }\right] $
must be a collection, too, where each $M_{i}\twoheadrightarrow _{\gamma
}N_{i}.$ By the induction hypothesis, $\Theta \left( M_{i}\right) \stackrel{D%
}{=}\Theta \left( N_{i}\right) $ for each $i\in 1..\left| M\right| .$ In
particular, for each $M_{i}$ there is an $N_{j}$ such that $M_{i}\equiv
N_{j}.$ The isomorphism follows by identifying each $i$ with the appropriate 
$j.$

This exhausts the cases and completes the proof.%
\endproof%

\subsection{Observational Semantics}

We provide another type of semantics for the $\lambda ^{p}$-calculus called
its \emph{observational semantics.} A formalism's observational semantics
expresses the computation as a whole:\ preparing the input, waiting for the
evaluation, and observing the result. The observational semantics relation
between $\lambda ^{p}$-terms and $\lambda $-terms is denoted$~\multimap $.
It is given by a single rule for the $\lambda ^{p}$-calculus. 
\begin{equation}
\frac{M\rightsquigarrow v\quad \Theta \left( v\right) =N}{M\multimap N}\text{%
(ObsP)}  \label{obs-p}
\end{equation}

\subsection{Examples}

A useful term of the $\lambda ^{p}$-calculus is a random number generator.
We would like to define a term that takes as input a Church numeral 
\underline{$n$} and computes a collection of numerals from \underline{$0$}
to \underline{$n$}. This can be represented by the following primitive
recursive $\lambda ^{p}$-term. 
\begin{equation}
\underline{\text{R}}\equiv \underline{\text{PRIM-REC}}\,\left( \lambda
k.\lambda p.\left( k,p\right) \right) \,\underline{\text{0}}
\end{equation}

\noindent Then for instance $\underline{\text{R}}\,\underline{\text{3}}%
=\left( 3,2,1,0\right) .$

The following term represents a random walk. Imagine a man that at each
moment can either walk forward one step or backwards one step. If he starts
at the point $0$, after $n$ steps, what is the distribution of his position? 
\begin{equation}
\underline{\text{W}}\equiv \underline{\text{PRIM-REC}}\,\left( \lambda
k.\lambda p.\left( \underline{\text{P}}p,\underline{\text{S}}p\right)
\right) \,\underline{\text{0}}
\end{equation}

\noindent We assume we have extended Church numerals to negative numbers as
well. This can be easily done by encoding it is a pair. We will show some of
the highlights of the evaluation of $\underline{\text{W}}\,\underline{\text{3%
}}.$ Note that $\underline{\text{W}}\,\underline{\text{1}}=\left( \underline{%
-1},\underline{1}\right) .$%
\begin{equation}
\begin{array}{lll}
\underline{\text{W}}\,\underline{\text{3}} & = & \underline{\text{P}}\left( 
\underline{\text{W}}\,\underline{\text{2}}\right) ,\underline{\text{S}}%
\left( \underline{\text{W}}\,\underline{\text{2}}\right) \\ 
& = & \underline{\text{P}}\left( \underline{\text{P}}\left( \underline{\text{%
W}}\,\underline{\text{1}}\right) ,\underline{\text{S}}\left( \underline{%
\text{W}}\,\underline{\text{1}}\right) \right) ,\underline{\text{S}}\left( 
\underline{\text{P}}\left( \underline{\text{W}}\,\underline{\text{1}}\right)
,\underline{\text{S}}\left( \underline{\text{W}}\,\underline{\text{1}}%
\right) \right) \\ 
& = & \underline{\text{P}}\left( \underline{\text{P}}\left( \underline{\text{%
W}}\,\underline{\text{1}}\right) ,\underline{\text{S}}\left( \underline{%
\text{W}}\,\underline{\text{1}}\right) \right) ,\underline{\text{S}}\left( 
\underline{\text{P}}\left( \underline{\text{W}}\,\underline{\text{1}}\right)
,\underline{\text{S}}\left( \underline{\text{W}}\,\underline{\text{1}}%
\right) \right) \\ 
& = & \underline{\text{P}}\left( \underline{\text{P}}\left( \underline{-1},%
\underline{1}\right) ,\underline{\text{S}}\left( \underline{-1},\underline{1}%
\right) \right) ,\underline{\text{S}}\left( \underline{\text{P}}\left( 
\underline{-1},\underline{1}\right) ,\underline{\text{S}}\left( \underline{-1%
},\underline{1}\right) \right) \\ 
& = & \underline{\text{P}}\left( \left( \underline{-2},\underline{0}\right)
,\left( \underline{0},\underline{2}\right) \right) ,\underline{\text{S}}%
\left( \left( \underline{-2},\underline{0}\right) ,\left( \underline{0},%
\underline{2}\right) \right) \\ 
& = & \left( \left( \underline{-3},\underline{-1}\right) ,\left( \underline{%
-1},\underline{1}\right) \right) ,\left( \left( \underline{-1},\underline{1}%
\right) ,\left( \underline{1},\underline{3}\right) \right) \\ 
& \equiv & \left( \underline{-3},\underline{-1},\underline{-1},\underline{1},%
\underline{-1},\underline{1},\underline{1},\underline{3}\right)
\end{array}
\end{equation}

\noindent Observing $\underline{\text{W}}\,\underline{\text{3}}$ yields $%
\underline{-1}$ with probability $\frac{3}{8},$ $\underline{1}$ with
probability $\frac{3}{8},$ $\underline{-3}$ with probability $\frac{1}{8}$,
and $\underline{3}$ with probability $\frac{1}{8}.$

\section{The Lambda-Q Calculus}

The $\lambda ^{q}$-calculus is an extension of the $\lambda ^{p}$-calculus
that allows easy expression of \emph{quantumized }algorithms. A quantumized
algorithm differs from a randomized algorithm in allowing negative
probabilities and in the way we sample from the resulting distribution.

Variables and abstractions in the $\lambda ^{q}$-calculus have \emph{phase}.
The phase is nothing more than a plus or minus sign, but since the result of
a quantumized algorithm is a distribution of terms with phase, we call such
a distribution by the special name \emph{superposition}. The major
difference between a superposition and a distribution is the observation
procedure. Before randomly picking an element, a superposition is
transformed into a distribution by the following two-step process. First,
all terms in the superposition that are identical except with opposite phase
are cancelled. They are both simply removed from the superposition. Second,
the phases are stripped to produce a distribution. Then, an element is
chosen from the distribution randomly, as in the $\lambda ^{p}$-calculus.

The words \emph{phase} and \emph{superposition} come from quantum physics.
An electron is in a superposition if it can be in multiple possible states.
Although the phases of the quantum states may be any angle from $0{{}^{\circ
}}$ to $360{{}^{\circ }}$, we only consider binary phases. Because we use
solely binary phases, we will use the words \emph{sign} and \emph{phase }%
interchangeably in the sequel.

A major disadvantage of the $\lambda ^{p}$-calculus is that it is impossible
to compress a collection. Every reduction step at best keeps the collection
the same size. Quantumized algorithms expressed in the $\lambda ^{q}$%
-calculus, on the other hand, can do this as easily as randomized algorithms
can generate random numbers. That is, $\lambda ^{q}$-terms can contain
subterms with opposite signs which will be removed during the observation
process.

\subsection{Syntax}

The following grammar describes the $\lambda ^{q}$-calculus. 
\begin{equation}
\begin{tabular}{|ll|}
\hline
$
\begin{array}{ll}
S & \in \text{\emph{Sign}} \\ 
x & \in \text{\emph{Variable}} \\ 
M & \in \text{\emph{LambdaQTerm}} \\ 
w & \in \text{\emph{WffQ}}
\end{array}
$ & $
\begin{array}{l}
\text{Sign, or phase} \\ 
\text{Variables} \\ 
\text{Terms of the }\lambda ^{q}\text{-calculus} \\ 
\text{Well-formed formulas of the }\lambda ^{q}\text{-calculus}
\end{array}
$ \\ 
&  \\ 
$
\begin{array}{lll}
S & ::= & + \\ 
& \,\,\,| & - \\ 
&  &  \\ 
M & ::= & Sx \\ 
& \,\,\,| & M_{1}M_{2} \\ 
& \,\,\,| & S\lambda x.M \\ 
& \,\,\,| & M_{1},M_{2} \\ 
&  &  \\ 
w & ::= & M_{1}=M_{2}
\end{array}
$ & $
\begin{array}{l}
\text{positive} \\ 
\text{negative} \\ 
\\ 
\text{signed variable} \\ 
\text{application} \\ 
\text{signed abstraction} \\ 
\text{collection} \\ 
\\ 
\text{well-formed formula}
\end{array}
$ \\ \hline
\end{tabular}
\newline
\label{lambda-q syntax}
\end{equation}

Terms of the $\lambda ^{q}$-calculus differ from terms of the $\lambda ^{p}$%
-calculus only in that variables and abstractions are \emph{signed}, that
is, they are preceded by either a plus (+)\ or a minus (-)\ sign. Just as $%
\lambda $-terms could be read as $\lambda ^{p}$-terms, we would like $%
\lambda ^{p}$-terms to be readable as $\lambda ^{q}$-terms. However, $%
\lambda ^{p}$-terms are unsigned and cannot be recognized by this grammar.

Therefore, as is traditionally done with integers, we will omit the positive
sign. An unsigned term in the $\lambda ^{q}$-calculus is abbreviatory for
the same term with a positive sign. With this convention, $\lambda ^{p}$%
-terms can be seen as $\lambda ^{q}$-terms all of whose signs are positive.
Also, so as not to confuse a negative sign with subtraction, we will write
it with a logical negation sign\ ($\lnot $). With these two conventions, the 
$\lambda ^{q}$-term $+\lambda x.+x-\!x$ is written simply $\lambda x.x\lnot
x.$

Instead of these conventions, we could just as well have rewritten the
grammar of signs so that the positive sign was spelled with the empty string
(traditionally denoted by the Greek letter $\epsilon $) and the negative
signs was spelled with the logical negation sign. We would have gotten the
alternative grammar below. 
\begin{equation}
\begin{array}{lll}
S^{\prime } & ::= & \epsilon \\ 
& \,\,\,| & \lnot
\end{array}
\label{alternative grammar for signs}
\end{equation}

\noindent However, this would have suggested an asymmetry between positive
and negative signs and allowed the interpretation that negatively signed
terms are a ``type'' of positively signed terms. On the contrary, we want to
emphasize that there are two distinct kinds of terms, positive and negative,
and neither is better than the other. There is no good reason why $\lambda
^{p}$-terms should be translated into positively signed $\lambda ^{q}$-terms
and not negatively signed ones. This arbitrariness is captured better as a
convention than a definition.

Finally, we adhere to the same parenthesization and precedence rules as the $%
\lambda ^{p}$-calculus. In particular, we continue the use of the
abbreviatory notation $\left[ M_{i}^{i\in S}\right] $ for collections of
terms, although we will not recast the parenthesization and ordering
invariance theorem (\ref{theorem:order/paren invariance}) for terms of the $%
\lambda ^{q}$-calculus. The modifications to the proof are mild.

\subsection{Syntactic Identities}

We want to give a name to the relationship between two terms that differ
only in sign.

\begin{definition}
\label{defn: opposite}A $\lambda ^{q}$-term $M$ is the \emph{opposite }of a $%
\lambda ^{q}$-term $N$, written $M\equiv \overline{N},$ if either 
\[
M\equiv S_{1}x\text{ and }N\equiv S_{2}x 
\]

\noindent where $S_{1}$ and $S_{2}$ are different signs, or 
\[
M\equiv S_{1}\lambda x.M^{\prime }\text{ and }N\equiv S_{2}\lambda
x.N^{\prime } 
\]

\noindent \noindent where $S_{1}$ and $S_{2}$ are different signs, and $%
M^{\prime }\equiv N^{\prime }.$
\end{definition}

\noindent Note that not all terms have opposites but if $M\equiv \overline{N}
$ then it follows that $N\equiv \overline{M}.$

We define substitution of terms in the $\lambda ^{q}$-calculus as a
modification of substitution of terms in the $\lambda ^{p}$-calculus. We
rewrite the seven rules of the $\lambda ^{p}$-calculus to take account of
the signs of the terms. First, we introduce the function notated by sign
concatenation, defined by the following four rules: 
\begin{eqnarray}
++ &\mapsto &+ \\
+- &\mapsto &- \\
-+ &\mapsto &- \\
-- &\mapsto &+
\end{eqnarray}

\noindent Notating this in our alternative syntax for signs (\ref
{alternative grammar for signs}), these rules can be summarized by the
single rewrite rule 
\begin{equation}
\lnot \lnot \mapsto \epsilon
\end{equation}

\noindent because the concatenation of a sign $S$ with $\epsilon $ is just $%
S $ again. Now we can use this function in the following substitution rules. 
\begin{equation}
\begin{array}{ll}
1.\;\left( Sx\right) \left[ N/x\right] \equiv SN & \text{for variables }y\not%
{\equiv}x \\ 
2.\;\left( Sy\right) \left[ N/x\right] \equiv Sy &  \\ 
3.\;\left( PQ\right) \left[ N/x\right] \equiv \left( P\left[ N/x\right]
\right) \left( Q\left[ N/x\right] \right) &  \\ 
4.\;\left( S\lambda x.P\right) \left[ N/x\right] \equiv S\lambda x.P &  \\ 
5.\;\left( S\lambda y.P\right) \left[ N/x\right] \equiv S\lambda y.\left(
P\left[ N/x\right] \right) & \text{if }y\not{\equiv}x\text{ and }y\notin
FV\left( N\right) \\ 
6.\;\left( S\lambda y.P\right) \left[ N/x\right] \equiv S\lambda z.\left(
P\left[ z/y\right] \left[ N/x\right] \right) & \text{if }y\not{\equiv}x\text{
and }y\in FV\left( N\right) \\ 
& \text{and }z\notin FV(P)\bigcup FV\left( N\right) \\ 
7.\;\left( P,Q\right) \left[ N/x\right] \equiv \left( P\left[ N/x\right]
,Q\left[ N/x\right] \right) & 
\end{array}
\label{substitution-q}
\end{equation}

Where did we use the sign concatenation function in the above substitution
rules?\ It is hidden in rule (1). Consider $\left( \lnot x\right) \left[
\lnot \lambda y.y/x\right] \equiv \lnot \lnot \lambda y.y.$ This is not a $%
\lambda ^{q}$-term by grammar (\ref{lambda-q syntax}). Applying the sign
concatenation function yields $\lambda y.y,$ which is a $\lambda ^{q}$-term.
However, we could not have rewritten rule (1) to be explicit about the sign
of $N$ because $N$ may be an application or a collection and therefore not
have a sign.

\subsection{Reduction}

The $\gamma $-relation of the $\lambda ^{q}$-calculus is of the same form as
that of the $\lambda ^{p}$-calculus.

\begin{equation}
\gamma ^{q}\triangleq \left\{ 
\begin{array}{l}
\left( \left[ M_{i}^{i\in 1..m}\right] \left[ N_{j}^{j\in 1..n}\right]
,\left[ M_{i}^{i\in 1..m}N_{j}^{j\in 1..n}\right] \right) \\ 
\text{such that }M_{i},N_{j}\in LambdaQTerm,\,m>1\text{ or }n>1
\end{array}
\right\}  \label{gamma-q}
\end{equation}
We omit the superscript when it is clear from context if the terms under
consideration are $\lambda ^{p}$-terms or $\lambda ^{q}$-terms. We still
write $\gamma \left( M\right) $ for the $\gamma $-normal form of $M.$
Theorems (\ref{thm: gamma-p CR}) and (\ref{thm: gamma-p NF exists}) are
easily extendible to terms of the $\lambda ^{q}$-calculus so $\gamma \left(
M\right) $ is well-defined.

We extend the $\beta $-relation to deal properly with signs.

\begin{equation}
\beta ^{q}\triangleq \left\{ 
\begin{array}{l}
\left( \left( S\lambda x.M\right) N,SM\left[ N/x\right] \right) \\ 
\text{such that }S\in \text{\emph{Sign}},S\lambda x.M\text{ and }N\in
LambdaQTerm
\end{array}
\right\}
\end{equation}

\noindent We refer to $\beta $-reduction for the $\lambda $-calculus, the $%
\lambda ^{p}$-calculus, and the $\lambda ^{q}$-calculus all with the same
notation when there is no risk of ambiguity.

\subsection{Evaluation Semantics}

We modify the call-by-value evaluation semantics of the $\lambda ^{p}$%
-calculus.

There are three rules for the call-by-value evaluation semantics of the $%
\lambda ^{q}$-calculus. 
\begin{eqnarray*}
&&\frac {}{v\rightsquigarrow v}\text{(Refl)\qquad \qquad (for }v\text{ a
value)} \\
&&\frac{\gamma \left( M\right) \rightsquigarrow S\lambda x.P\quad \gamma
\left( N\right) \rightsquigarrow N^{\prime }\quad \gamma \left( SP\left[
N^{\prime }/x\right] \right) \rightsquigarrow v}{MN\rightsquigarrow v}\text{%
(Eval)} \\
&&\frac{\gamma \left( M\right) \rightsquigarrow v_{1}\quad \gamma \left(
N\right) \rightsquigarrow v_{2}}{\left( M,N\right) \rightsquigarrow \left(
v_{1},v_{2}\right) }\text{(Coll)}
\end{eqnarray*}

\subsection{Observation}

\label{lambda-q observation}We define an observation function $\Xi $ from $%
\lambda ^{q}$-terms to $\lambda $-terms as the composition of a function $%
\Delta $ from $\lambda ^{q}$-terms to $\lambda ^{p}$-terms with the
observation function $\Theta $ from $\lambda ^{p}$-terms to $\lambda $-terms
defined in (\ref{lambda-p observation}). Thus, $\Xi =\Theta \circ \Delta $
where we define $\Delta $ as follows. 
\begin{eqnarray}
\Delta \left( Sx\right) &=&x \\
\Delta \left( S\lambda x.M\right) &=&\lambda x.\Delta \left( M\right) \\
\Delta \left( M_{1}M_{2}\right) &=&\Delta \left( M_{1}\right) \Delta \left(
M_{2}\right) \\
\Delta \left( M\equiv \left[ M_{i}^{i\in 1..\left| M\right| }\right] \right)
&=&\left[ \Delta \left( M_{i}^{i\in \left\{ i\,\,\,|\,\,\,M_{i}\not{\equiv}%
\overline{M_{j}}\,\,\text{for }j\in 1..\left| M\right| \right\} }\right)
\right]  \label{Collection case for delta}
\end{eqnarray}

\noindent The key is in the case (\ref{Collection case for delta}) where the
argument to $\Delta $ is a collection. In this case, the function $\Delta $
does two things. First, it removes those pairs of terms in the collection
that are opposite. Then, it recursively applies itself to each of the
remaining terms.

Note that unlike the observation function $\Theta $ of the $\lambda ^{p}$%
-calculus, the observation function $\Xi $ of the $\lambda ^{q}$-calculus is
not total. For some $\lambda ^{q}$-term $M$, $\Xi \left( M\right) $ does not
yield a $\lambda $-term. An example of such a term is $M\equiv x,\lnot x$
because $\Delta \left( M\right) $ is the collection $M$ with all pairs of
opposites removed. However, the empty collection is not a $\lambda ^{p}$%
-term. Therefore, some $\lambda ^{q}$-terms cannot be observed. The
non-totality of the observation function $\Xi $ does not limit the $\lambda
^{q}$-calculus because careful programming can always insert a unique term
into a collection prior to observation to ensure observability. There is
thus no need to add distinguished tokens to the $\lambda ^{q}$-calculus such
as \underline{\emph{error}} or \underline{\emph{unobservable}}.

Because $\Xi =\Theta \circ \Delta ,$ the definition of statistical
indistinguishability (\ref{defn: statistically indistinguishable}) applies
to $\Xi \left( M\right) $ and $\Xi \left( N\right) $ as well, if both $%
\Delta \left( M\right) $ and $\Delta \left( N\right) $ exist. Although
observing a $\lambda ^{p}$-term is statistically indistinguishable from
observing its $\gamma $-normal form, observing a $\lambda ^{q}$-term is, in
general, statistically distinguishable from observing its $\gamma $-normal
form.

\subsection{Observational Semantics}

The observational semantics for the $\lambda ^{q}$-calculus is similar to
that of the $\lambda ^{p}$-calculus (\ref{obs-p}). It is given by a single
rule. 
\begin{equation}
\frac{M\rightsquigarrow v\quad \Xi \left( v\right) =N}{M\multimap N}\text{%
(ObsQ)}  \label{obs-q}
\end{equation}

\subsection{Examples}

We provide one example. We show how satisfiability may be solved in the $%
\lambda ^{q}$-calculus. We assume possible solutions are encoded some way in
the $\lambda ^{q}$-calculus and there is a term $\underline{\text{CHECK}_{f}}
$ that checks if the fixed Boolean formula $f$ is satisfied by a particular
truth assignment, given as the argument. The output from this is a
collection of $\underline{\text{T}}$ (truth) and $\underline{\text{F}}$
(falsity) terms. We now present a term that will effectively remove all of
the $\underline{\text{F}}$ terms. It is an instance of a more general
method. 
\begin{equation}
\underline{\text{REMOVE-F}}\equiv \lambda x.\,\underline{\text{IF}}%
\,x\,x\,\left( x,\lnot x\right)
\end{equation}

We give an example evaluation. 
\begin{equation}
\begin{array}{lll}
\underline{\text{REMOVE-F}}\,\left( \underline{\text{F}},\underline{\text{T}}%
,\underline{\text{F}}\right) & \equiv & \left( \lambda x.\,\underline{\text{%
IF}}\,x\,x\,\left( x,\lnot x\right) \right) \left( \underline{\text{F}},%
\underline{\text{T}},\underline{\text{F}}\right) \\ 
& \twoheadrightarrow _{\gamma } & \left( 
\begin{array}{l}
\left( \lambda x.\,\underline{\text{IF}}\,x\,x\,\left( x,\lnot x\right)
\right) \underline{\text{F}}, \\ 
\left( \lambda x.\,\underline{\text{IF}}\,x\,x\,\left( x,\lnot x\right)
\right) \underline{\text{T}}, \\ 
\left( \lambda x.\,\underline{\text{IF}}\,x\,x\,\left( x,\lnot x\right)
\right) \underline{\text{F}}
\end{array}
\right) \\ 
& \twoheadrightarrow _{\beta } & \left( \left( \underline{\text{F}},\lnot 
\underline{\text{F}}\right) ,\underline{\text{T}},\left( \underline{\text{F}}%
,\lnot \underline{\text{F}}\right) \right) \\ 
& \equiv & \left( \underline{\text{F}},\lnot \underline{\text{F}},\underline{%
\text{T}},\underline{\text{F}},\lnot \underline{\text{F}}\right)
\end{array}
\end{equation}

\noindent Observing the final term will always yield $\underline{\text{T}}.$
Note that the drawback to this method is that if $f$ is unsatisfiable then
the term will be unobservable. Therefore, when we insert a distinguished
term into the collection to make it observable, we risk observing that term
instead of $\underline{\text{T}}.$ At worst, however, we would have a
fifty-fifty chance of error.

Specifically, consider what happens when the argument to $\underline{\text{%
REMOVE-F}}$ is a collection of $\underline{\text{F}}^{\prime }$s$.$ Then $%
\underline{\text{REMOVE-F}}\,\underline{\text{F}}=\left( \underline{\text{F}}%
,\lnot \underline{\text{F}}\right) .$We insert $\underline{\text{I}}\equiv
\lambda x.x$ which, if we observe, we take to mean that either $f$ is
unsatisfiable or we have bad luck. Thus, we observe the term $\left( 
\underline{\text{I}},\underline{\text{F}},\lnot \underline{\text{F}}\right)
. $ This will always yield $\underline{\text{I}}.$ However, we cannot
conclude that $f$ is unsatisfiable because, in the worst case, the term may
have been $\left( \underline{\text{I}},\underline{\text{REMOVE-F}}\,%
\underline{\text{T}}\right) =\left( \underline{\text{I}},\underline{\text{T}}%
\right) $ and we may have observed $\underline{\text{I}}$ even though $f$
was satisfiable. We may recalculate until we are certain to an arbitrary
significance that $f$ is not satisfiable.

Therefore, applying $\underline{\text{REMOVE-F}}$ to the results of $%
\underline{\text{CHECK}_{f}}$ and then observing the result will yield $%
\underline{\text{T}}$ only if $f$ is satisfiable.

\section{Conclusion}

We have seen two new formalisms. The $\lambda ^{p}$-calculus allows
expression of randomized algorithms. The $\lambda ^{q}$-calculus allows
expression of quantumized algorithms. In these calculi, observation is made
explicit, and terms are presumed to exist in some Heisenberg world of \emph{%
potentia}.

This work represents a new direction of research. Just as the $\lambda $%
-calculus found many uses, the $\lambda ^{p}$-calculus and the $\lambda ^{q}$%
-calculus may help discussion of quantum computation in the following ways.

\begin{enumerate}
\item  Quantum programming languages can be specified in terms of the $%
\lambda ^{q}$-calculus and compared against each other.

\item  Algorithms can be explored in the $\lambda ^{q}$-calculus on a higher
level than quantum Turing machines, which, like classical Turing machines,
are difficult to program.

\item  An exploration of the relationship between the $\lambda ^{q}$%
-calculus and quantum Turing machines, quantum computational networks, or
other proposed quantum hardware, may provide insights into both fields.
\end{enumerate}

We have seen some algorithms for the $\lambda ^{p}$-calculus and the $%
\lambda ^{q}$-calculus. It should not be difficult to see that the $\lambda
^{p}$-calculus can simulate a probabilistic Turing machine and that the $%
\lambda ^{q}$-calculus can simulate a quantum Turing machine. It should also
follow that a probabilistic Turing machine can simulate the $\lambda ^{p}$%
-calculus, with the exponential slowdown that comes from computing in the
world of reality rather than the world of \emph{potentia.} However, it is
not obvious that a quantum Turing machine can simulate the $\lambda ^{q}$
-calculus. An answer to this question, whether positive or negative, will be
interesting. If quantum computers can simulate the $\lambda ^{q}$-calculus
efficiently, then the $\lambda ^{q}$-calculus can be used as a programming
language directly. As a byproduct, satisfiability will be efficiently
solvable. If quantum computers cannot simulate the $\lambda ^{q}$-calculus
efficiently, knowing what the barrier is may allow the formulation of
another type of computer that can simulate it.


\begin{thebibliography}{9}
\bibitem{abadi-cardelli}  Abadi, Mart\'{i}n and Luca Cardelli. \emph{A
Theory of Objects}. Monographs in Computer Science, David Gries and Fred B.
Schneider editors. Springer-Verlag, New York: 1996. Chapter 6.

\bibitem{barendregt}  Barendregt, Hendrik Pieter. \emph{The lambda calculus:
its syntax and semantics.} North-Holland, New York: 1981.

\bibitem{church}  Church, Alonzo. \emph{An unsolvable problem of elementary
number theory.} American Journal of Mathematics 58, 1936, pp.345-363.

\bibitem{R4RS}  \emph{Revised}$^{4}$\emph{\ Report on the Algorithmic
Language Scheme.} William Clinger and Jonathan Rees editors. November 2,
1991.

\bibitem{deutsch 85}  Deutsch, David. \emph{Quantum theory, the
Church-Turing principle and the universal quantum computer}. Proc. R. Soc.
Lond. A400, 1985, pp.97-117.

\bibitem{deutsch 89}  Deutsch, David. \emph{Quantum computational networks}.
Proc. R. Soc. Lond. A425, 1989, pp.73-90.

\bibitem{deutsch 92}  Deutsch, David. \emph{Quantum computation}. Physics
World, June, 1992, pp.57-61.

\bibitem{heisenberg}  Heisenberg, Werner. \emph{Physics and philosophy. }%
Harper \&\ Brothers, New York: 1958.

\bibitem{simon}  Simon, Daniel. \emph{On the power of quantum computation}.
Proceedings of the 35th Annual Symposium on the Foundations of Computer
Science. IEEE Computer Society Press, Los Alamitos, CA: 1994.
\end{thebibliography}
\end{document}